\documentclass{PoS}

\usepackage{lscape}
\usepackage{afterpage}
\usepackage{amsmath}
\usepackage[square,comma,numbers,sort&compress]{natbib}
\usepackage{float}

\newcommand*{\no}{\noindent}
\newcommand*{\bea}{\begin{eqnarray}}
\newcommand*{\eea}{\end{eqnarray}}
\newcommand*{\be}{\begin{equation}}
\newcommand*{\ee}{\end{equation}}

\newcommand*{\pref}[1]{(\ref{#1})}

\newcommand*{\nn}{\nonumber}

\newcommand{\bma}{\begin{pmatrix}}
\newcommand{\ema}{\end{pmatrix}}

\title{A G$_2$-QCD neutron star}

\ShortTitle{A G$_2$-QCD neutron star}

\author{\speaker{Ouraman Hajizadeh}\thanks{Supported by the FWF DK W1203-N16}\\
        E-mail: \email{ouraman.hajizadeh@uni-graz.at}}

\author{Axel Maas\\
        E-mail: \email{axel.maas@uni-graz.at}\\\\
        Institute of Physics, NAWI Graz, University of Graz, Universit\"atsplatz 5, 8010 Graz, Austria}

\abstract{The determination of the properties of neutron stars from the underlying theory, QCD, is still an unsolved problem. This is mainly due to the difficulty to obtain reliable results for the equation of state for cold, dense QCD. As an alternative route to obtain qualitative insights, we determine the structure of a neutron star for a modified version of QCD: By replacing the gauge group SU(3) with the exceptional Lie group G$_2$, it is possible to perform lattice simulations at finite density, while still retaining neutrons. Here, results of these lattice simulations are used to determine the mass-radius relation of a neutron star for this theory. The results show that phase changes express themselves in this relation. Also, the radius of the most massive neutron stars is found to vary very little, which would make radius determinations much simpler if this would also be true in QCD.}

\FullConference{34th annual International Symposium on Lattice Field Theory\\
		24-30 July 2016\\
		University of Southampton, UK}

\begin{document}

\section{Introduction}

The properties of neutron stars \cite{Glendenning:1997wn,Steiner:2010fz,Kapusta:2006pm} have been the subject of astronomical studies and observations since decades. Nevertheless, a satisfying explanation of their properties is still out of reach. The main reason for this is that it was not yet possible to derive the equation of state governing neutron stars from the fundamental theory, i.\ e.\ QCD, at finite density and small or zero temperature \cite{Friman:2011zz}. The reason is that lattice gauge theory, the mainstay of non-perturbative QCD calculations, suffers from the sign problem \cite{Gattringer:2010zz,deForcrand:2010ys}. Alternative methods are either still not sufficiently far developed or make heavy use of modeling, which reduces their predictivity \cite{Leupold:2011zz,Buballa:2003qv,Pawlowski:2010ht,Braun:2011pp}.

Turning this around, data of neutron stars could provide us with a laboratory for strong interactions in this cold and dense regime of the phase diagram \cite {Lattimer:2006xb,Glendenning:1997wn,Steiner:2010fz}. Hence, they make it possible to observe large-scale effects of non-Abelian gauge theories. Also the recent discovery of gravitational waves \cite{Abbott:2016blz} will provide data on the bulk properties, giving further constraints \cite{DelPozzo:2013ala}.

Pending a solution for QCD, it would be highly interesting to know whether any generic signatures of a realistic equation of state of a non-Abelian gauge theory can manifest in the features of neutron stars. To make progress, the avenue exploited here will be to study a QCD-like theory without sign problem. To have any chance at qualitative similarity requires this theory to have fermionic baryons, especially neutrons, and have rather similar features as QCD itself. A theory with these feature is G$_2$-QCD, i.\ e.\ QCD where the gauge group SU(3) is replaced by the exceptional Lie group G$_2$ \cite{Holland:2003jy,Maas:2012wr}, see \cite{Maas:2012ts} for a review of the properties of this theory. A rough outline of the full phase diagram \cite{Maas:2012wr}, detailed spectroscopical data and properties of the structure at zero temperature and finite baryon density \cite{Wellegehausen:2013cya} are available for this theory from lattice simulations.
 
Based on these results, we determine the mass-radius relation of a G$_2$-QCD neutron star. Because the data are for one flavor and comparatively heavy pions, it will probably not be a semi-quantitatively good guideline, but some qualitative features can be expected, and seem to be, rather robust. This will be discussed in more detail elsewhere \cite{Hajizadeh:unpublished}. There, we will also consider the impact of varying parameters like the pion mass. In total, the features obtained are not too different from the expectations from models  \cite{Glendenning:1997wn,Steiner:2010fz,Kapusta:2006pm}, which is by itself promising. Most interesting is that we find evidence for different phases inside this model neutron star to surface in astronomically observable quantities. 
 
\section{G$_2$-QCD}

We will briefly rehearse here the relevant features of G$_2$-QCD for our investigation \cite{Wellegehausen:2013cya}. G$_2$ is a subgroup of SO(7), and has 7 quark colors and 14 gluons \cite{Holland:2003jy}, and all representations are real. Therefore it does not suffer from a sign problem \cite{Maas:2012wr,Kogut:2000ek}.

Furthermore, the theory shares many features with QCD. Among those similarities are the coincidence of chiral and deconfinement transition in the quenched case \cite{Danzer:2008bk}. The spectrum is rich \cite{Wellegehausen:2013cya}, and contains fermionic baryons \cite{Holland:2003jy,Wellegehausen:2013cya}, in particular three-quark states, to be called neutrons in the following. At sufficiently light quark masses it is expected that these are the lightest fermionic baryons \cite{Maas:2012wr,Wellegehausen:2013cya}. At the same time, it is possible to construct color-neutral states with any number of quarks and varying numbers of gluons \cite{Holland:2003jy,Wellegehausen:2013cya}, as well as mesons and glueballs.

Due to the reality of the representations there is an enlarged chiral symmetry, the Pauli-G\"ursey symmetry \cite{Kogut:2000ek,Wellegehausen:2013cya}. Thus, non-trivial symmetry breaking is possible for the one-flavor case considered here. The Goldstone bosons in this case are diquarks \cite{Wellegehausen:2013cya}.

In the following, we use as input the lattice data for the light ensemble of \cite{Wellegehausen:2013cya}, i.\ e.\ a Goldstone mass of 247 MeV with a neutron of 938 MeV mass fixing the scale. For details of the simulations we refer also to this work.
 
\section{Mass-Radius relation of neutron stars}     

The most pertinent feature of neutron stars is their mass-radius relation \cite{Glendenning:1997wn}. To obtain it, we follow the approximation of Tolman, Oppenheimer and Volkoff: We assume a static and spherically symmetric metric, and an energy-momentum tensor of an ideal fluid, which are realistic and simplifying assumptions. This yields the Tolman-Oppenheimer-Volkoff (TOV) equation  \cite{Glendenning:1997wn}
\begin{equation}
\frac{dp(r)}{dr} = - \frac{[p(r)+\epsilon(r)]~[M(r)+4\pi r^3p(r)]}{r~[r-2M(r)]}\nn,
\end{equation}
\no where \(M(r)\) is the mass-energy enclosed within the radius \(r\):
\begin{equation}
M(r) = 4 \pi \int_0^r \epsilon(r') ~ r'^2 ~ dr'\label{mass},
\end{equation}
\no and $p$ and $\epsilon$ are the pressure and energy density, respectively. Thus, knowledge of both quantities are required to solve the TOV equation.

For this purpose, we will employ two different versions of the pressure and energy density. As a baseline, we employ the well-known \cite{Glendenning:1997wn} case of non-interacting neutrons. This helps us understanding the impact of interactions. 

\begin{figure}
\begin{center}
\includegraphics[width=8cm,type=pdf,ext=.pdf,read=.pdf]{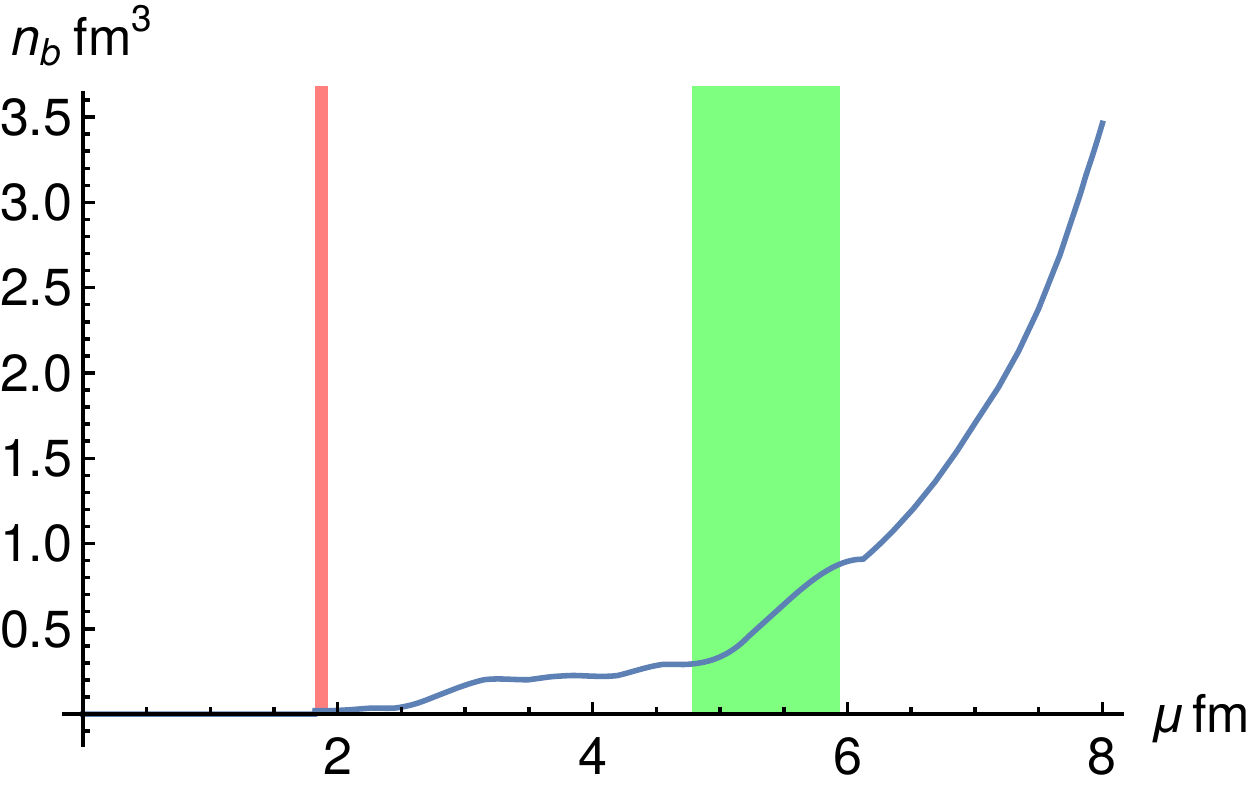} 
\end{center}
\caption{\label{fig:n}The employed baryon density as a function of baryo-chemical potential, interpolated from the lattice results of \cite{Wellegehausen:2013cya}. The red band shows the mass of the lightest particle in the spectrum, the diquark Goldstone, while the green band is the mass of the neutron, both values also from \cite{Wellegehausen:2013cya}, both including statistical errors only.} 
\end{figure}

For the full interacting theory, we neglect contributions from the kinetic energy. As the rest mass is much larger than the average thermal energy, this should be a decent approximation. Therefore, the energy density is entirely given by the density of neutrons as a function of the baryo-chemical potential $\mu$, $n(\mu)$ and their rest mass $m_n$, $\epsilon=m_n n(\mu)$, and the pressure is then \cite{Kapusta:2006pm}
\begin{equation}
  p= n\mu-\epsilon=(\mu-m_n)n(\mu)\nn.
\end{equation}
\no Thus, if the baryo-chemical potential drops to the mass of the neutron, the pressure drops to zero, and the surface of the neutron star is reached. The baryon density is taken from the lattice simulations of G$_2$-QCD of \cite{Wellegehausen:2013cya}, and depicted in figure \ref{fig:n}. Note that the lattice only yields discrete points, which we have interpolated.

The actual baryon density already starts to differ from zero at a baryo-chemical potential below the neutron mass, as the Goldstones are diquarks and also carry baryon number \cite{Wellegehausen:2013cya}. Here, we assume that the neutron star consists entirely of neutrons, i.\ e.\ the diquarks, because of their bosonic nature, do not contribute significantly in the chemical composition. Relaxing this assumption will be studied elsewhere \cite{Hajizadeh:unpublished}, but so far does not seem to have strong qualitative impacts. A further problem is that the mesh of points in baryo-chemical potential on the lattice does not include the point $\mu=m_n$. Around this chemical potential, the pressure must be the one of ordinary matter to provide a stable neutron star \cite{Glendenning:1997wn}, which we enforce in the fit. This yields a systematic, but slight, quantitative dependence on where in the chemical potential this fit is enforced, but this has also little qualitative impact. This will also be studied elsewhere in more detail \cite{Hajizadeh:unpublished}.

Finally, rewriting the TOV equation using this pressure only and in terms of the baryo-chemical potential yields
\begin{equation}
\frac{d\mu(r)}{dr} = - \frac{[\mu(r)]~[M(r)+4\pi r^3p(r)]}{r~[r-2M(r)]}\label{tov}.
\end{equation}
\no To obtain the mass-radius relation we solve the coupled equations \pref{mass} and \pref{tov} for various values of the chemical potential at the core of the neutron star, starting from slightly above $m_n$. The initial conditions are obtained from the requirement of zero mass at the center of the neutron stars and the corresponding pressure value. The solution itself is possible using standard algorithms for ordinary differential equations.

\section{Results}

As figure \ref{fig:n} shows, the neutron star will sample only the region starting from the green band, due to our assumption of the chemical composition. The most important feature in the baryon density is a step-like structure around and above this point, which could be associated with a phase change \cite{Wellegehausen:2013cya}. This is interpreted as the onset of a phase where neutrons or heavier fermionic hadrons dominate \cite{Wellegehausen:2013cya}, and which is thus precisely the relevant phase here. Of course, at very high chemical potentials lattice artifacts will play a role \cite{Maas:2012wr,Wellegehausen:2013cya}, but it turns out that this is at a value of chemical potential where the neutron star is no longer stable.

\begin{figure}
 \begin{center}
   \includegraphics[width=0.9\textwidth,type=pdf,ext=.pdf,read=.pdf]{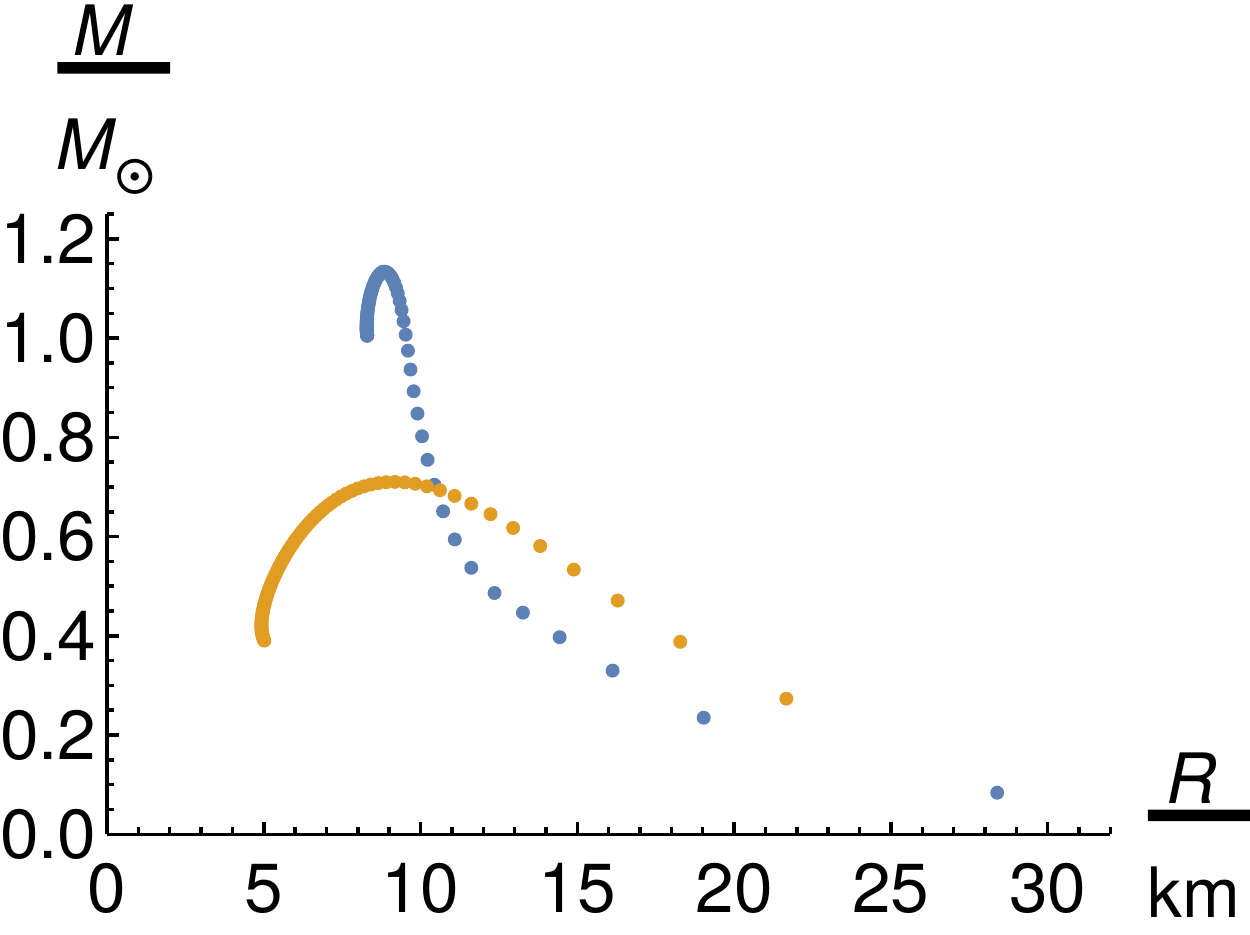}
 \end{center}
 \caption{\label{fig:mr}Mass-radius relation for the G$_2$ neutron star (blue) in comparison to the result for the free Fermi gas (orange).}
\end{figure}

The results for the mass-radius-relation are shown in figure \ref{fig:mr}. What is immediately seen from the plot is that the interaction increases the maximum mass of the neutron star and also that it shows less stability in response to perturbations in the radius, comparing to the free case. Turned around, given an (almost) maximum mass neutron star, its radius is essentially fixed. The end-point of the mass-radius curve is at a chemical potential where lattice artifacts start to become an issue, but are not yet dominating.

At large radius the free neutron gas allows larger masses, than the interacting case, which leads to a steeper rise. Conversely, at a smaller radius, about 11 km, the rise for the interacting case becomes much quicker. This can be traced back to the baryon density shown in figure \ref{fig:n}. The corresponding central chemical potential coincides with a region slightly above the masses of ground state neutrons and $\Delta$s. Starting from this point, the equation of state appears to enter a new region dominated by fermionic hadrons \cite{Wellegehausen:2013cya}. Thus, this rapid change of properties impresses itself on the mass-radius relation. This could imply that changes of slope of the mass-radius relation of neutron stars could also be related to different phases.

\begin{figure}
\includegraphics[width=0.58\textwidth,type=pdf,ext=.pdf,read=.pdf]{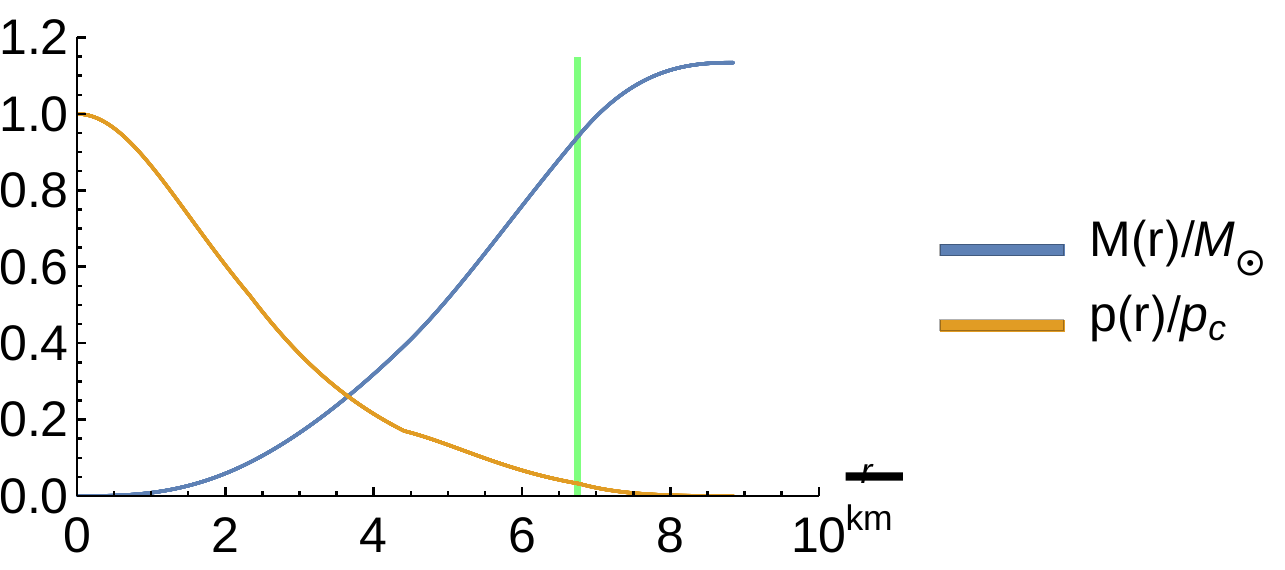}\includegraphics[width=0.42\textwidth,type=pdf,ext=.pdf,read=.pdf]{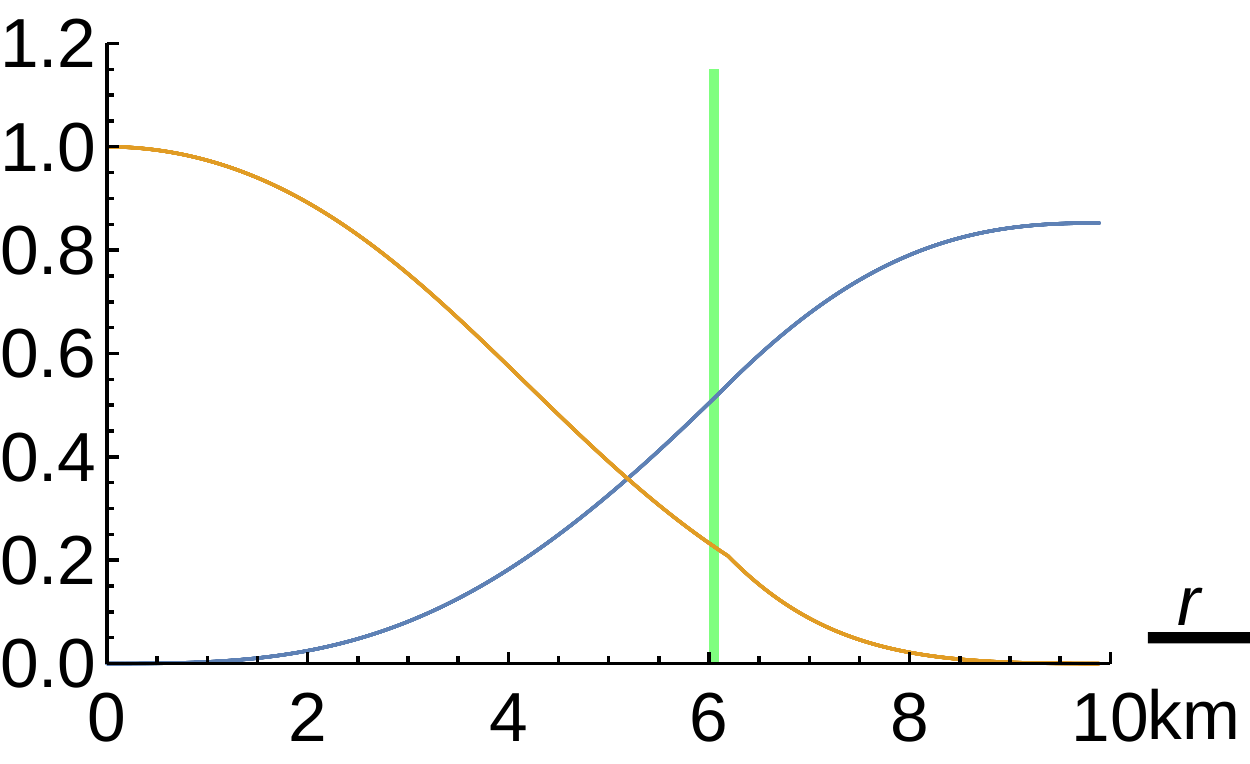}
\caption{\label{fig:profile} Accumulated mass (blue) and pressure (orange) as a function of distance from the center of the neutron star for two different central chemical potential, $\mu_c=7.6$ fm$^{-1}$ (left) and $\mu_c=6$  fm$^{-1}$ (right). The green line indicates the last step before the neutron threshold in the baryon density in figure \protect\ref{fig:n}.}
\end{figure}

It is also interesting to look inside the G$_2$ neutron star to see, whether in their profile the different phases are distinguished. This is studied in figure \ref{fig:profile}. In fact, the change is seen also in the pressure profile , though the effect is much less pronounced than in the density itself. However, the mass distribution is not visibly affected.

\begin{figure}
\includegraphics[width=0.5\textwidth,type=pdf,ext=.pdf,read=.pdf]{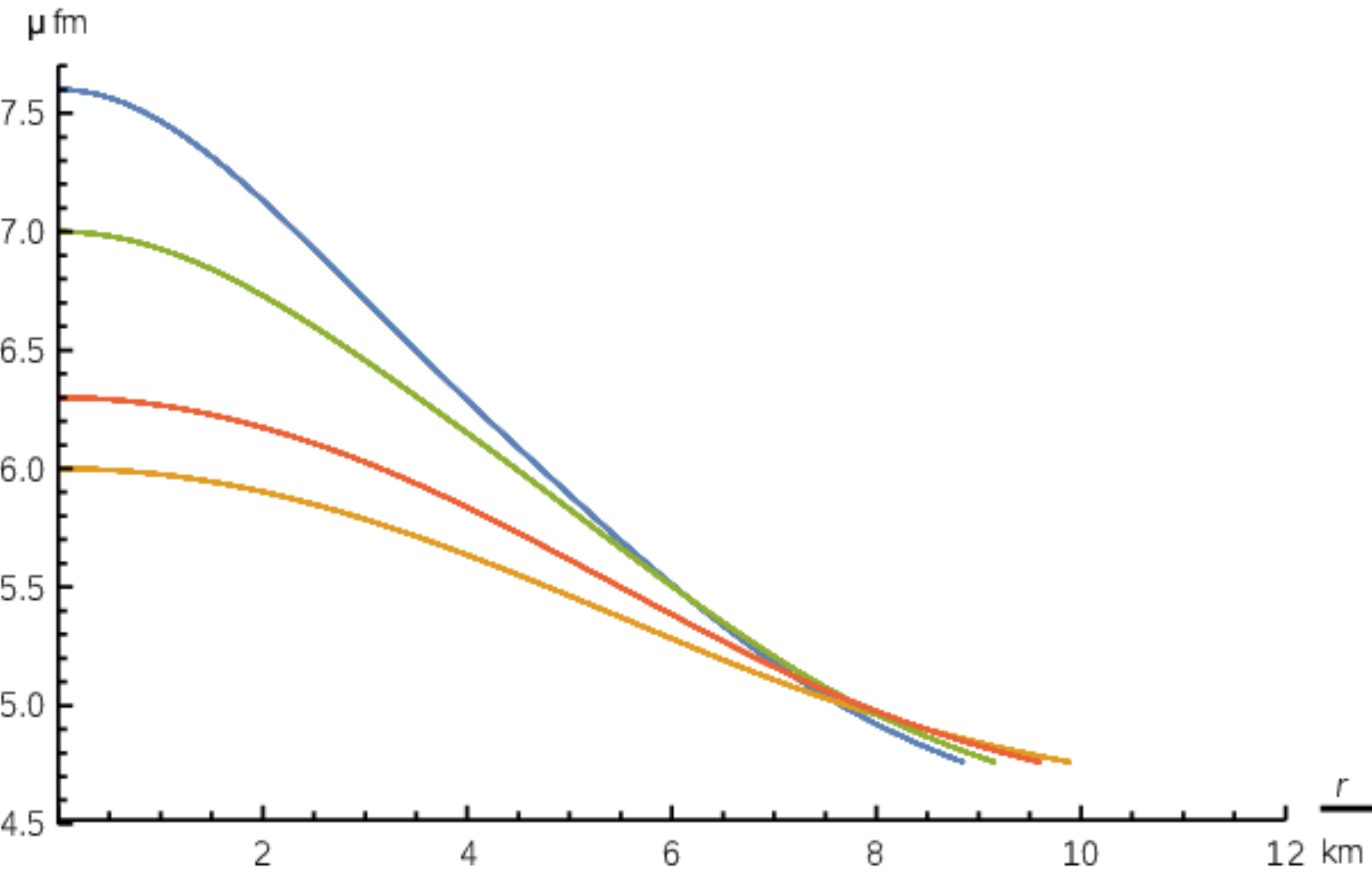}\includegraphics[width=0.5\textwidth,type=pdf,ext=.pdf,read=.pdf]{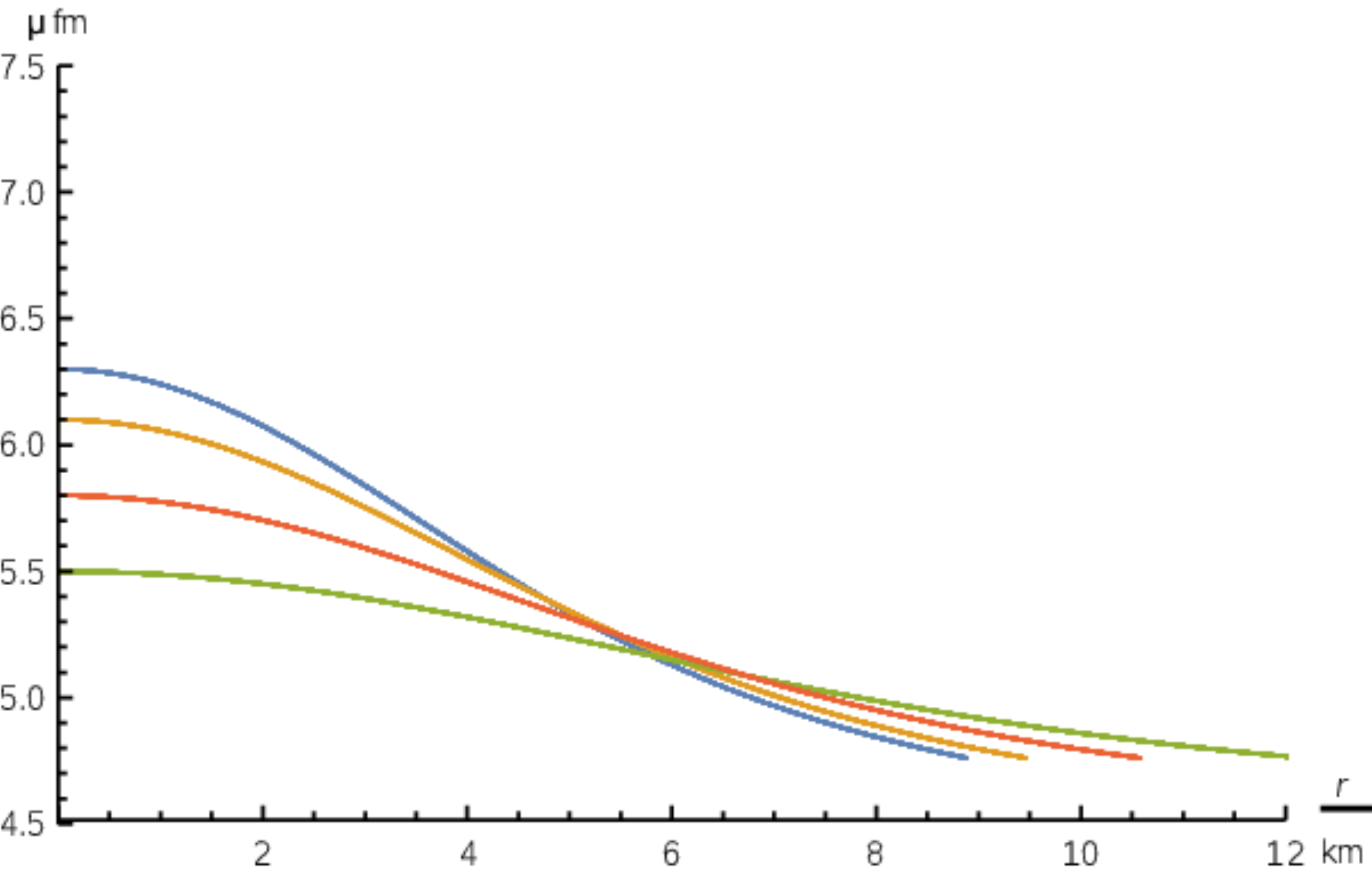}
\caption{\label{fig:mu}The chemical potential as a function of distance from the center for different values of the central chemical potential for G$_2$-QCD (left) and for the free neutron gas (right)}
\end{figure}

The different profiles also explain why the radius distribution around the maximum mass is much narrower than in the free case. In figure \ref{fig:mu} the profiles of $\mu(r)$ curves  do not diverge considerably after the crossing point, in comparison to the free gas, while they are quickly changing in the interacting gas. Also, a much higher central chemical potential is supported than for the free gas, leading to the larger mass. It is also visible how the slope is negative throughout, thus supporting the stability of the star.

\section{Conclusions}

We have provided, for the first time, a mass-radius relation for a neutron star based on the equation of state of a QCD-like theory with fermionic baryons, using lattice data as an input.

The results show a number of expected features. In particular, the curves become steeper and the maximum mass increases, without exceeding the maximum value possible from general relativity. Of course, given the fact that the theory is different from QCD and that only one flavor with a rather heavy Goldstone is included, the increase to roughly 1.2 solar masses instead of the more than two possible in QCD is not entirely unexpected.

However, even this could already hint that either the masses of the quarks or the number of degrees of freedom play a significant role for the maximum neutron star mass. Less speculative and more visible in the data is that the phase change seen in the baryon density impresses itself into the mass-radius relation of the neutron star. It also survives in the pressure gradient. This implies that already observing the mass-radius relation in sufficient detail could resolve the old question of whether a neutron star is a, more or less, monolithic object, or whether it contains a multitude of different phases.

These kind of qualitative insights show how important the study of QCD-like theories is for the understanding of possible features of the real case. Of course, the results here could be improved in various ways, some of which we will investigate in \cite{Hajizadeh:unpublished}, while others will have to await improved results for the input equation of state. Still, this gives eventually the hope that we could learn how to interpret qualitative features of neutron stars from such model systems.

\bibliographystyle{bibstyle}
\bibliography{bib}

\end{document}